Tortuosity Entropy: a measure of spatial complexity of behavioral changes in animal movement data

Xiaofeng Liu[*], Ning Xu, Aimin Jiang

([1]College of IoT Engineering, Hohai University，Changzhou，213022, China)

([2]Changzhou key laboratory of robotics and intelligent technology，Changzhou，213022，China)

Correspondence should be addressed to: xfliubme@gmail.com

The goal of animal movement analysis is to understand how organisms explore and exploit complex and varying environments. Animals usually exhibit varied and complicated movements, from apparently deterministic behaviors to highly random ones. This is critical for assessing movement efficiency and strategies that are used to quantify and analyze movement trajectories. Here we introduce a tortuosity entropy (TorEn) based on comparison of parameters (e.g. heading, bearing, speed, of consecutive points) in movement trajectory, which is a simple measure for quantifying the behavioral change in animal movement data at a fine scale. In our approach, the differences between pairwise successive track points are transformed inot symbolic sequences, then we map these symbols into a group of pattern vectors and calculate theie information entropy. Tortuosity entropy can be easily applied to arbitrary real-world data, whether deterministic or stochastic, stationary or non-stationary. We test the algorithm on both simulated trajectories and real trajectories and show that both mixed segments in synthetic data and different phases in real movement data are identified accurately. The results show that the algorithm is applicable to a variety of situations, indicating that our approach is a promising tool to reveal the behavioral pattern in movement data.

## 1. Introduction

The movement of an organism, the displacement of individual organisms in space over time, plays a fundamental role in the structure and dynamics of populations (Nathan, Getz et al. 2008). As pointed out in a recent paper (Nathan, Getz et al. 2008), movement research has shifted from quantifying population redistribution to quantifying movement of individuals. In recent years, the study of movement of organisms has been a rapidly expanding field of research, producing a growing body of work devoted to movement of individual organisms due to technological advances of tracking devices, e.g., GPS and Argos (Tomkiewicz, Fuller et al. 2010). Full range technological advances from the microelectronics, wireless telecommunication, satellite telecommunication, to high-capacity storage make possible the development of a high performance track logger that is miniaturized size and ultra light weight. As a result, there has been an accumulation of movement data of various organisms (Nathan, Getz et al. 2008), which raises a challenging analyses of movement data, including quantitatively charactering the spatiotemporal trajectories and retrieving relevant information to reveal a causal link between animal movements and internal states and external factors.

A variety of methods for extracting information about behavior or for characterizing behavioral change in movement data have been developed and applied (Batschelet 1981; Bovet and Benhamou 1988; Dicke and Burrough 1988; Nams 1996; Fauchald and Tveraa 2003; Benhamou 2004; Roberts, Guilford et al. 2004; Horne, Garton et al. 2007; Barraquand and Benhamou 2008; Gurarie, Andrews et al. 2009; Gaucherel 2011). Straightness index, defined as the ratio of the beeline distance between the start and end of a trajectory to the length of a travelled trajectory, is a the simplest but most straightforward measure of how behavior varies over spatiotemporal scales (Batschelet 1981), and has been further extended into a multi-scale straightness index in a recent study (Postlethwaite, Brown et al. 2013) . Sinuosity is an index of the tortuosity of a random search trajectory that is determined by both distribution of changes of direction and by step length (Bovet and Benhamou 1988). Fractal dimension is another metric used to characterize the tortuosity of a trajectory (Dicke and Burrough 1988; Nams 1996; Atkinson, Rhodes et al. 2002; Fritz, Said et al. 2003). These indices are a kind of measures of the spatial features of trajectories. Some methods are widely applied methods to quantify the temporal information in movement data. First-passage time (FPT) (Fauchald and Tveraa 2003), defined as the time required for an animal to cross a given area, is a scale dependent metric which indicates search effort of animal along a path (Fauchald and Tveraa 2003; Pinaud 2008). Similar to first-passage time, residence time (Barraquand and Benhamou 2008) in the vicinity of successive path locations is used to identify the profitable place. Furthermore, the periodicity of movement recursion has received attention. Well-established Fourier and wavelet analyses are used to extract periodic patterns or to detect the regimes shift (Polansky, Wittemyer et al. 2010; Gaucherel 2011; Riotte-Lambert, Benhamou et al. 2013). Unlike the traditional feature extraction methods, Roberts and his colleagues have proposed positional entropy (Roberts, Guilford et al. 2004) as an index of navigational trajectory uncertainty by means of calculating the stochastic complexity of path (Rezek and Roberts 1998). They argued that it is better to measure the tortuosity of the tracks using the stochastic complexity (Roberts, Guilford et al. 2004).

Methods based on concepts of entropy and complexity are popular and have been widely applied to the studies of time series. Animal movements can be described as a time series of movement steps. They usually exhibit behavioral heterogeneity showing that an animal often behaves in diverse modes, such as travelling, foraging, resting and escaping predators during the trajectory. For example, animal's foraging movement consists of extensive and intensive search phases that are driven by environment characteristics.

In this paper, we focus on the pattern of local movement change by presenting an algorithm of quantifying behavioral change in a small scale. We compare the two successive track points, denoted by a symbol, then map these symbol series into a pattern vector. Finally, we calculate the information entropy of pattern vector variable, termed tortuosity entropy. Tortuosity entropy can be used to various parameters of trajectory, heading, speed, bearing, and position. Moreover, it can be easily applied to arbitrary real-world data - deterministic or stochastic, stationary or non-stationary.

The remainder of this paper is organized as follows: Section 2 addresses our algorithms. Section 3 shows the performance of TorEn on synthetic trajectories. Section 4 demonstrates its applications to real GPS movement data; Section 5 includes the discussion and future works. Finally, we conclude our paper in section 6.

## 2. Tortuosity Entropy

In this section, we define the tortuosity entropy and introduce the basic idea. The procedure of algorithm consists of two main steps and is summarized in Fig. 1. The basic idea is that the relative change of consecutive points in a movement trajectory is quantified and denoted as a series of symbols, and the frequency of each specific symbol is counted. Then information entropy of symbolic sequences is defined as tortuous entropy, which is an indicator of the extent of local change in animal movement data. The larger tortuosity entropy is, the more tortuous and the wigglier the trajectory is. The procedure of tortuosity entropy consists of two main steps: (1) quantification and symbolization of the movement path, (2) calculation of information entropy, described below in detail. To simplify our calculation, we assume that each track point in movement data is recorded at a fixed time interval. We consider a raw movement trajectory that consists of $n$ samples, $\{x_i\} = (x_1, x_2, \cdots, x_{N-1}, x_N)$, $for\ t = 1, \cdots N$, which can be one kind of movement parameters, e.g. absolute position, heading, or speed.

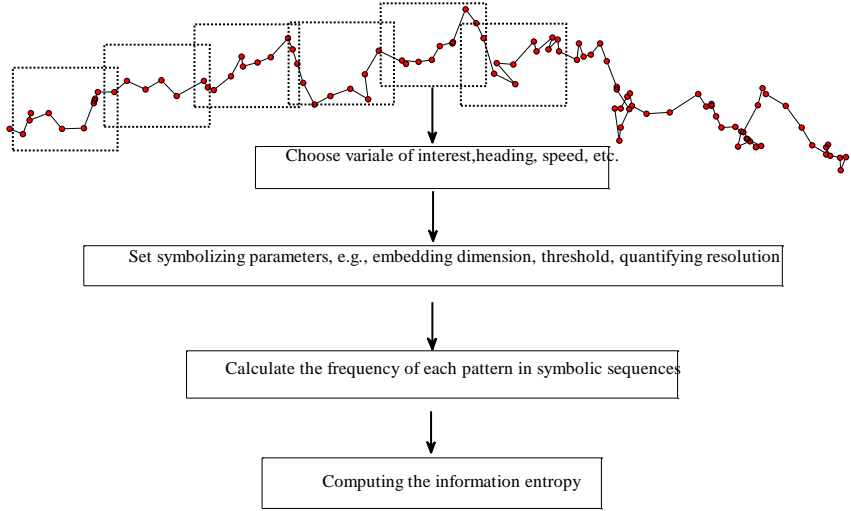

Fig. 1. Schematic description of the steps of the tortuosity entropy.

## 2.1 Quantification and Symbolization

We are interested in the change of consecutive samples. There exist a variety of approaches to quantify a time series. The quantifying resolution is an integer $\alpha \geq 2$. For coarse quantification, we set quantifying resolution equal to 2, thus the trajectory $\{x_i\}$ are simply transformed in terms of the ups and downs into the series $\{s_i\}$, $i = 1, \cdots, N-1$, made up of two symbols, 0 for going up and 1 for going down, as described in Eq. (1).

$$s_i = \begin{cases} 0, & x_{i+1} \geq x_i \\ 1, & else \end{cases} \tag{1}$$

While set $\alpha$ equal to 3, the $\{x_i\}$ can be easily quantified in terms of Eq. (2).

$$s_i = \begin{cases} 0, & |x_{i+1} - x_i| \leq \lambda \\ 1, & x_{i+1} > x_i + \lambda \\ 2, & x_{i+1} < x_i - \lambda \end{cases} \tag{2}$$

where $\lambda$ is a quantification threshold.

## 2.2 Calculation of information entropy of symbolic sequences

Entropy is a measure of uncertainty in a random variable. In information theory, entropy is a measure of information content, which is defined as the average uncertainty in a random variable. Larger entropy means higher uncertainty.

For $\{s_i\}$, $i = 1, \cdots, K$, we get $K - m + 1$ pattern vectors $V(1), \cdots, V(i), \cdots, V(K - m + 1)$ through the overlapped embedding procedure where $V(i) = [s(i), s(i + 1), ..., x(i + m - 1)]$, $i = 1, ..., K - m + 1$, when we set the embedding dimension parameter $m$.

The relative frequency of any unique pattern vector, $p_j$, is the amount of any unique pattern vector, denoted as $q_j$, that appears in the time series divided by the total number of pattern vectors,

$$p_j = \frac{q_j}{K - m - 1}. \quad (3)$$

The tortuosity entropy of order $m \geq 2$ is defined as

$$H(m) = -\sum_{j=1}^{a^m} p_j \ln p_j \quad (4)$$

Obviously, $H(m)$ is bounded in $[0, m \cdot \ln(\alpha)]$ where the lower bound is reached for the time series with an approximate constant change (e.g., an ordered sequence values) and the upper bound for a completely random system. In general, $H(m)$ is normalized by $(m - 1)$,

$$h_m = H(m) \big/ (m - 1) \quad (5)$$

Tortuosity entropy is calculated for different embedding dimensions $m$. For practical purposes, we recommend $m = 3, 4, 5$.

## 3. Test on simulated trajectory data

We first simulate an oriented travel trajectory with a consistent bias in a given preferred direction $\gamma$, which is modeled using biased random walks (BRWs) for a long time. The trajectory that an animal travels towards a given target (e.g. homing or migration) is typical an oriented path that contains a consistent bias in the preferred direction. We generate an oriented trajectory using the following algorithm. The path begins to evolve at the point $(0,0)$ with an initial direction randomly taken from $0 \sim 360°$ and an initial step randomly taken from $0.8 \sim 1.2$.

$$\begin{cases} x_{i+1} = x_i + l_i \cos\theta_i \\ y_{i+1} = y_i + l_i \sin\theta_i \end{cases} \quad (6)$$

$$\theta_i = \arctan(\alpha \sin(\gamma + \xi) + (1 - \alpha)\sin\theta_{i-1}, \alpha\cos(\gamma + \xi) + (1 - \alpha)\cos\theta_{i-1}) \quad (7)$$

where the length of each step, $l_i$, is picked from a normal distribution with mean $l_{i-1}$ and variance $\sigma^2$, $\alpha$ is a weight factor within $(0 \sim 1)$ and $\xi$ is an angular random variable picked from a Von Mises distribution (Benhamou and Bovet 1992).

The formulas (7) means that the direction, $\theta_i$, taken at each step can be computed as the weighted sum of preferred direction, $\gamma$, and the previous movement direction, $\theta_{i-1}$.

The three different synthetic trajectories and their dynamical tortuosity entropy are shown in Fig. 2. These trajectories are simulated using different turning angles picked from a Von Mises distribution with different concentration. In terms of the Von Mises distribution, these trajectory are characterized by different tortuosity. Parts of each trajectory are also shown in enlarged view (see Fig. 2a-c). We calculate the dynamical TorEn on bearing of the trajectory, and shown the results in Fig. 2d. It is clear that TorEn is highest for the trajectory shown in Fig. 2c, followed by that in Fig. 2b. TorEn is the smallest in Fig. 2a. Clearly, TorEn is an effective indicator of the tortuosity of three simulated trajectories, and moreover, dynamical TorEn captures the non-stationarity of tortuosity of each simulated trajectory.

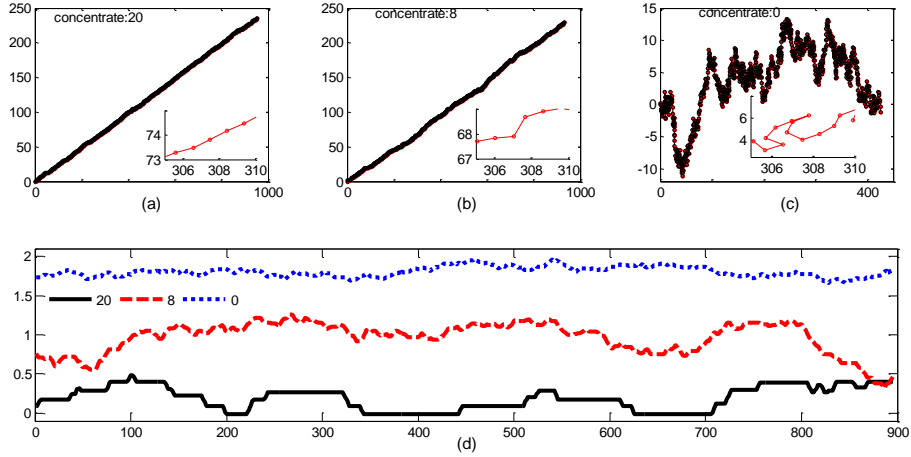

Fig. 2. Illustration of measure of tortuosity using TorEn. Upper panel: three simulated trajectories in terms of formulas (6) . The turning angle is picked from Von Mises distribution with location $\mu = \pi / 12$ and concentration $\kappa = 20$ in plot (a), with location $\mu = \pi / 12$ and concentration $\kappa = 8$ in plot (b), with location $\mu = \pi / 12$ and concentration $\kappa = 0$ in plot (c) . Lower panel: the dynamical TorEn of three trajectories, with threshold $\theta = 30^\circ$, window size $M = 100$, and embedding dimension $m = 3$.

### 3.2 Analyzing composite paths and detecting behavioral change

Trajectories travelling for a long time are more likely to encompass multiple movement patterns, due to the animal's response to external environment changes (e.g. landscape characters) and internal state changes. The animal's movements tend to be relatively straight movement paths when the animal migrates or disperses and take more tortuous paths when it forages or breeds. Therefore, it is necessary and fundamental to segment the trajectory and to detect the behavioral change in order to understand the behavioral mechanisms in the animal movement.

We consider a mixed trajectory that is suppoee to be mdae up of two situations. One is based on an oriented travel while the other is based on an area-restricted search, both of which are simply synthetic using formulas (6). Slightly different from the above simulation where the direction is computed using formulas (7), the direction of each step, $\theta_i$, is picked from a Von Mise distribution with mean $\theta_{i-1}$ and concentration $\kappa$. We generate the two segments with the same parameters as a previous study (Postlethwaite, Brown et al. 2013). In view of the fact that an area-restricted search is characterized by slow speed, initial speed is slightly slower for random search paths than for straight paths. Fig. 3a shows an example of a composite trajectory with 1,000 track points, including two segments of an approximately straight path ($\kappa = 1000, \sigma = 0.01, l_1 = 1$) with 200 track points interspersed with three segments of random search ($\kappa = 0, \sigma = 0.01, l_1 = 0.8$) with 200 track points. The initial direction of either the straight path or the random search is picked from a uniform distribution.

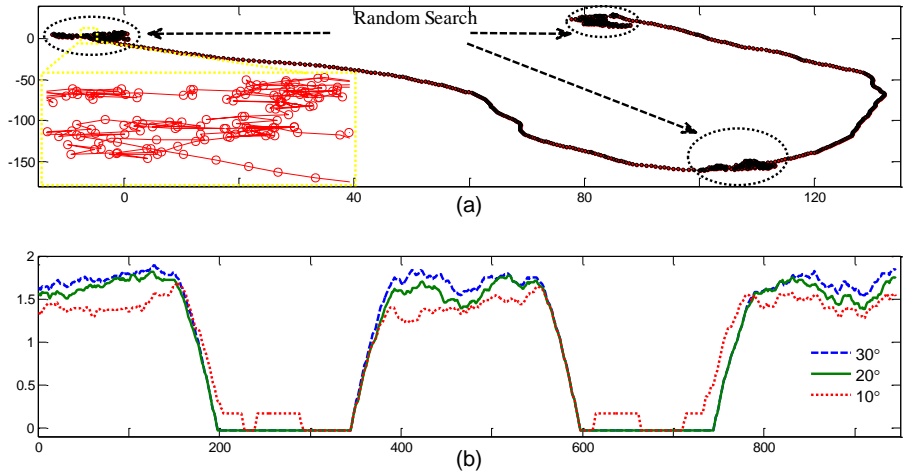

Fig.3. Illustration of detection of different segments in synthetic trajectory by means of TorEn. In the upper panel, three random search phases are interspersed with two straight travelling where each segment includes 200 track points. In the lower panel, dynamical TorEn of three trajectories with window size $M = 50$ and embedding dimension $m = 3$. To compare the effect of threshold on TorEn, results for three threshold parameter are superimposed in the lower panel.

Tortuosity entropy is shown in Fig. 3b. It is very clear in Fig. 2b that transitions between two behavior patterns exit in trajectory. TorEn for straight segments is extremely slow (equal to zero) which is consistent with observations presented in Fig. 1d. In contrast, TorEn is much higher for search segments than for straight segments. Taking the transition width due to the window size of dynamical TorEn (herein M = 50 track points), the duration of TorEn for each segment is the same as the length of each segment. This indicates that TorEn is capable of tracking the transition from one pattern to another one. As shown in Fig. 2b, the TorEn calculated with threshold $\theta = 30^\circ$ is almost identical to that with $\theta = 20^\circ$. However, both of them are slightly different from that with $\theta = 10^\circ$, in particular, for straight segments and transition phases, although the overall trends are approximate. This suggests thresholds that are too small make TorEn excessively sensitive to the fluctuations in trajectory, which may introduce a little distortion in dynamical TorEn.

Area-restricted search that is characterized by reduced speeds and frequent turns is obviously distinct from the approximately straight line. It is relatively easy to identify two behaviors in terms of their tortuosity entropy. In order to demonstrate the performance of tortuosity entropy for detecting the behavioral change in composite random search in which many animals exhibit when looking for resources, we generate a composite random search interspersed with extensive foraging and intensive foraging phases in response to a resource rich area or a resource poor area, respectively. The step length is drawn from the a Weibull distribution rather than from a normal distribution, in view of the fact that a Weibull distribution has been used to model animal speed in previous studies (Morales, Haydon et al. 2004; Haydon, Morales et al. 2008; Sippel, Holdsworth et al. 2011). Directions of each step are modeled as Von Mises distributions. We use the formulas (6) to generate the synthetic trajectory. The parameter kappa in a Von Mises distribution is taken as 10 for the intensive foraging pattern and as 0 for the extensive foraging pattern, and the shape parameter $k$ and scale parameter $\lambda$ of a Weibull distribution are taken as 1 and 15 for the intensive foraging phase and 1 and 1.5 for the extensive foraging phase. A synthetic composite random search trajectory, alternatively presentation of an extensive foraging pattern and an intensive foraging pattern, is shown in Fig 4a, where each pattern lasts for 200 track points. An enlarged view of a part of each pattern is also presented in Fig 4a. We can clearly see that the intensive foraging pattern is characterized by low speed and strong sinuosity, but the extensive foraging pattern is characterized by high speed and small sinuosity.

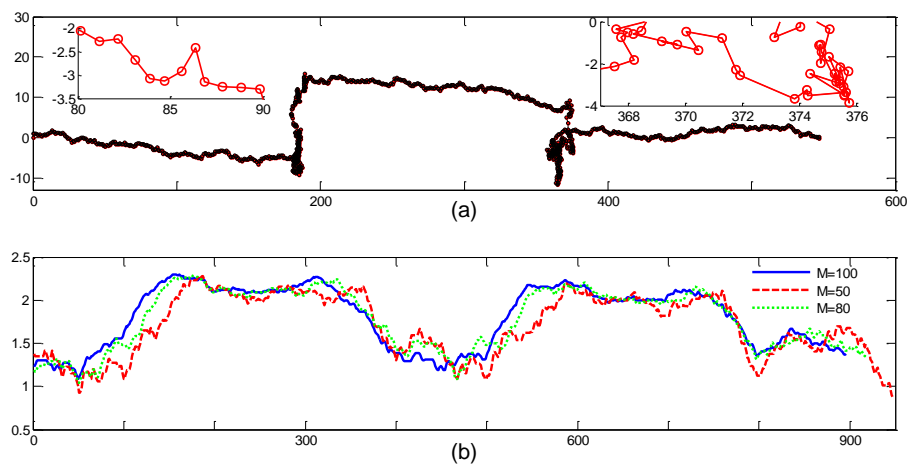

Fig.4. Illustration of detection of different phases in random search by means of TorEn. In the upper panel, a simulated trajectory is composed of three extensive search phases interspersed with two intensive search phases, in which each segment includes 200 track points. In the lower panel, the dynamical TorEn of three trajectories with embedding dimension $m = 3$ and threshold $\theta = 30^{\circ}$ is presented. To compare the effect of window size on TorEn, results for three window size are presented in the lower panel.

As shown in Fig. 3b, TorEn on bearing of trajectory is much higher for intensive search than for extensive search, which is consistent with the common observation that intensive search is more tortuous than extensive search. We can also clearly see the transitions between two behaviors. Although the complexity of extensive search trajectory is lower in comparison with intensive trajectory, TorEn of extensive search trajectory is far higher than a straight line. As a result, the transition in Fig. 3b is not as significant as that in Fig. 2b. We also demonstrate the effect of window size on dynamical TorEn. Comparing the results for three window size, dynamical TorEn with shorter window can track the transition more precisely. The choice of parameters for calculating TorEn will be further explained in the section 5.

## 4. Application to real recorded movement data

We apply the tortuosity entropy to the global positing system (GPS) data collected from Galapagos albatross (phoebastria irrorata) published in the movebank database (Dodge, Bohrer et al. 2013).

Twenty eight adult albatrosses are tracked throughout the entire breeding season from June to September 2008. These birds breed at on Isla Española, Punta Cevallos (1.39 °S, 89.62 °W) , Punta Suarez (1.38 °S, 89.75 °W), as well as Isla de la Plata (1.58 °S, 81.15 °W). The original tracking data and data logger are described in detail in the section of "case study" of the paper by Dodge and his colleagues (Dodge, Bohrer et al. 2013). A single track (#4267-84830990-152) of an albatross is shown in Fig 5a. In terms of the true position, the bird first flees from the initial site (1.3726 °S, 89.7402 °W) towards the foraging area (the first around 50 points), then search this area from around 50 to 190 track points where the bird reorientates frequently and the trajectory is much tortuous. After around 190 track points, the bird begins flying back towards Punta Suarez.

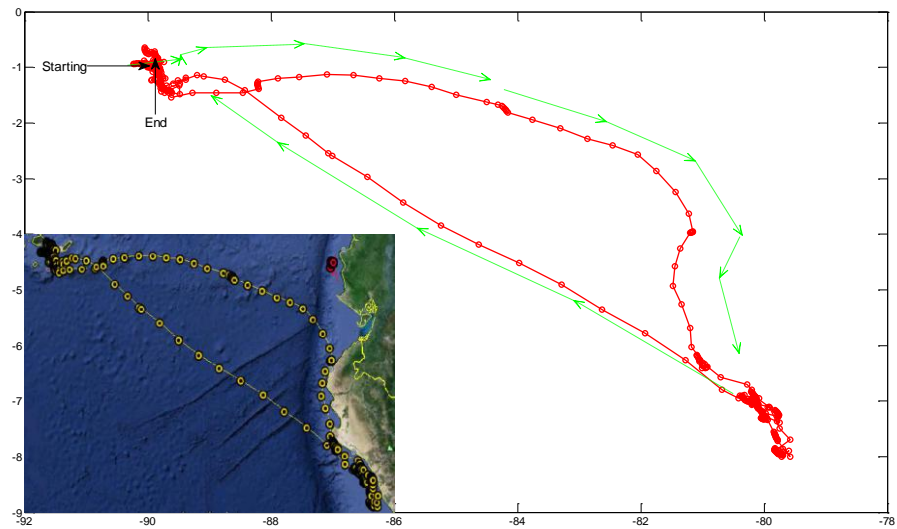

Fig. 5. GPS path of the albatross (#4267-84830990-152) to and fro Punta Suarez Island and the Peruvian coastal foraging area. The green dash line with arrows indicates the journey of albatross.

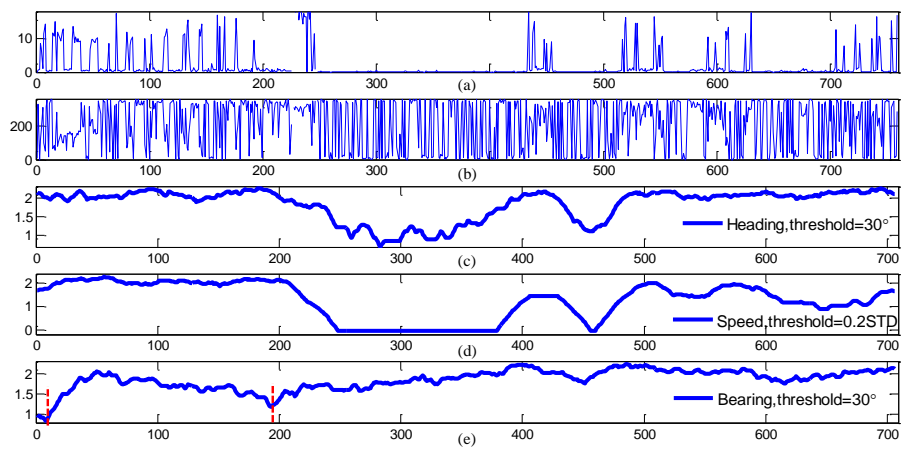

Fig. 6 Flight parameters speed (a) and heading (b) of trajectory in Fig. 5 and corresponding TorEn on heading (c),

speed (d) and bearing (e) respectively. Dynamical TorEn is calculated with M = 50, and m=3.

We show the basic parameters of flight, speed and heading, in Fig. 6 in order to explain the function of TorEn and its application to real movement data. Comparing speed curve in Fig. 6a and dynamical TorEn in Fig. 6d, we think overall TorEn can capture the change of speed. The trajectory is more tortuous in first 200 track points in Fig. 6d in terms of TorEn on speed, which is similar to TorEn on heading in Fig. 6c. This roughly indicates that the bird frequently changes its heading and speed with larger variation in the search stage. However, they are slightly different after about 500 track points. TorEn on heading is comparable to the initial 200 points; however, TorEn on speed is a little lower than that in the initial 200 points. This suggests that the bird exhibits different behavioral activity in these two stages.

Comparing the curve in Fig. 6e with that in Fig. 6c and in Fig. 6d, respectively, we find that TorEn on bearing offers other respects regarding animal movement. It can be clearly seen from the journey shown in Fig. 5 that the bird travels to and fro the coastal area and the island over water, from the beginning to around the 60[th] track point and from around the 226[th] to the 246[th] track points. Each travelling stage lasts for a very short period, about 60 track points forward to Peruvian coastal area and about 20 points backwards. When the albatross flies with high speed to and fro he coastal area and the island over water, its movement is an oriented travelling behavior. During the oriented-travelling stage the bearing of trajectory keeps stable towards the destination, thus the complexity of bearing should be relatively smaller than other stages, as illustrated in Fig. 3. We check TorEn on bearing and find that TorEn corresponding to travelling behavior is local minimum (as indicated by red dash line in Fig 6e). This suggests TorEn can captures accurately the two travelling stages, even the duration of one travel is far less than window size of dynamical entropy.

## 5. Discussion

In this paper, we present a new measure of tortuosity of animal movements based on symbolic entropy. This algorithm can be used to accurately identify the different segments involved in individual trajectory that are fundamental to the classification of diverse behavioral patterns. This approach is independent from assumptions about the goal of implement. Furthermore, TorEn can reveal all the aspects of animal movement characteristics because it is able to analyze the all parameters of trajectory, such as heading, bearing, speed, and various measures of the distance between successive track points. It is obviously distinct from traditional methods, e.g., straightness index, tortuosity, first passage time, and fractal dimension, which only offers a single static characteristic value of trajectory. It is different from the recent methods as well, e.g. multi-scale straightness index and positional entropy, which only exploit the positional change in trajectory but neglect the heading and speed changes. As an intrinsic advantage of working with symbols, TorEn possesses a higher efficiency of numerical computation. With this advantage, TorEn can be a promising tool for analyzing animal movements.

As mentioned above, in this study animal behavioral characteristics can be analyzed not only in relation to positions and direction in space, but also in relation to temporal or spatio-temporal parameters. Usually a parameter for analyzing the behavioral mechanism is chosen depending on assumptions of the movement. For example, turning angles is considered to be the most relevant parameter for investigating the search efficiency of non-oriented searches (Bartumeus, Catalan et al. 2008). The time spent in a limited area is deemed as a key parameter to quantify the search effort at pertinent spatial scales (Fauchald and Tveraa 2003; Pinaud 2008). Both of them are concentrated on one parameter of trajectory. In contrast, in a recent study which develop an approach for splitting and segmenting animal trajectories (Thiebault and Tremblay 2013), the authors consider behavior as a consequence of successive speed and direction choices, rather than just a single parameters, i.e., a set of positions in space. For our approach, we are very flexible in choosing various parameters of trajectory to study the behavioral mechanism underlying response to environmental change or landscape features, which makes TorEn capable of uncovering all the features of animal movements. This was demonstrated in the example in section 4.1, where TorEn is applied for analyzing the albatross' feeding and travelling behavior over water and islands. We can see the transition from travelling over water to other behaviors (e.g. feeding on an island) either from the speed and the heading (Fig. 5a and 5b) or from TorEn (Fig. 5c and 5d). In contrast, TorEn on the bearing of the albatross's trajectory effectively captures this transition between oriented travelling, searching or feeding, as shown in Fig. 5e.

Among the existing methods for extracting behavioral pattern or for identifying segments from animal movement data, most only offer a single summary value for a trajectory segment or even an entire trajectory. Wavelet transformation is a useful tool to expand the signal in a time-scale plane, and has been used to analyze the cyclicity of animal movements and their behaviors simultaneously over time and scale domain (Wittemyer, Polansky et al. 2008; Polansky, Wittemyer et al. 2010). Similarly, a new multi-scale measure, termed a 'multi-scale straightness index', extends the traditional straightness index into multi-scale (Postlethwaite, Brown et al. 2013), and is calculated repeatedly for sub-sections of individual trajectory, and thus, it offers a means of assessing behavior variability over time. In this respect, we are able to apply our method to the overlapped sliding window of animal trajectory as shown in Fig. 1, and thus we can get dynamical characteristics of behavioral variation of individual trajectories over time. We should point out that window size affects the temporal resolution of dynamical characteristics which is a common problem in sliding window approaches. Comparing the Fig. 3b and Fig. 4b, we can clearly see the effect of window size. Although it is more sensitive to abrupt changes in data for shorter window, we have to take a trade off between the temporal resolution and fluctuations. Empirically, a sliding window covers at least 50 track points for analyzing animal trajectory in small scale. For practical purposes, we recommend threshold $20^{\circ} \sim 30^{\circ}$ for analyzing the angle parameters.

Movement processes are the result of an animal's reaction to an internal state, an external stimulus, or complex interactions between an animal's internal state and external environmental stimuli. As a result, behavioral corresponds can exhibit multiple respects of movements, ranging from a suite of position in pace to various parameters, such as movement speed, heading, and bearing. These parameters represent different information about an animal's activities respectively. For example, when an albatross is feeding on an islands, it locomotes at extremely low speed (see Fig. 6a), but its heading still keep changing (see Fig. 6b). TorEn of speed is zero and TorEn of heading is also at the lowest period. On the contrary, TorEn of bearing is relatively higher, indicating that the trajectory direction is often varied. Thus, this pattern implies that an albatross could feed at situ, rather than at rest. This example shows how our method can fully capture the feature of an animal trajectory.

## 6. Concluding remarks

TorEn is based on basic symbolization procedures and symbolic entropy concepts, which makes it relatively robust to noise. Unlike the state space model, it is not able to estimate the behavioral modes TorEn can be applied flexibly on various parameters of trajectory as a post-hoc evaluation of movement data. Therefore, it is capable of revealing the holographic changes in individual movements, thus it is a powerful tool to identify the different segments in movements and to provide further understand of the behavioral strategies of a moving animal.

**Acknowledgements**

This research is supported in part by NSFC under the grant 60905060, 31101643, 61101158, by the Fundamental Research Funds for the Central Universities of China under grants 2011B11214, 2011B11114, 2012B07314, 2012B04014, and by Open Fund of Ministry of Education Key Lab of Broadband Wireless Communication and Sensor Network Technology under grant NYKL201305.

References

Atkinson, R. P. D., C. J. Rhodes, et al. (2002). "Scale-free dynamics in the movement patterns of jackals." Oikos **98**(1): 134-140.

Barraquand, F. and S. Benhamou (2008). "Animal movements in heterogeneous landscapes: identifying profitable places and homogeneous movement bouts." Ecology **89**(12): 3336-3348.

Bartumeus, F., J. Catalan, et al. (2008). "The influence of turning angles on the success of non-oriented animal searches." Journal of Theoretical Biology **252**(1): 43-55.

Batschelet, E. (1981). Circular statistics in biology. London, Acadmic Press.

Benhamou, S. (2004). "How to reliably estimate the tortuosity of an animal's path:: straightness, sinuosity, or fractal dimension?" Journal of Theoretical Biology **229**(2): 209-220.

Bovet, P. and S. Benhamou (1988). "Spatial analysis of animals' movements using a correlated random walk model." Journal of Theoretical Biology **131**(4): 419-433.

Dicke, M. and P. A. Burrough (1988). "Using fractal dimensions for characterizing tortuosity of animal trails." Physiological Entomology **13**(4): 393-398.

Dodge, S., G. Bohrer, et al. (2013). "The environmental-data automated track annotation (Env-DATA) system: linking animal tracks with environmental data." Movement Ecology **1**(1): 3.

Fauchald, P. and T. Tveraa (2003). "Using First-Passage Time in the Analysis of Area-Restricted Search and Habitat Selection." Ecology **84**(2): 282-288.

Fritz, H., S. Said, et al. (2003). "Scale–dependent hierarchical adjustments of movement patterns in a long–range foraging seabird." Proceedings of the Royal Society of London. Series B: Biological Sciences **270**(1520): 1143-1148.

Gaucherel, C. (2011). "Wavelet analysis to detect regime shifts in animal movement." Computational Ecology and Software **1**(2): 69-85.

Gurarie, E., R. D. Andrews, et al. (2009). "A novel method for identifying behavioural changes in animal movement data." Ecology Letters **12**(5): 395-408.

Haydon, D. T., J. M. Morales, et al. (2008). "Socially informed random walks: incorporating group dynamics into models of population spread and growth." Proceedings of the Royal Society B: Biological Sciences **275**(1638): 1101-1109.

Horne, J. S., E. O. Garton, et al. (2007). "ANALYZING ANIMAL MOVEMENTS USING BROWNIAN BRIDGES." Ecology **88**(9): 2354-2363.

Morales, J. M., D. T. Haydon, et al. (2004). "Extracting more out of relocation data: building movement models as mixtures of random walks." Ecology **85**(9): 2436-2445.

Nams, V. O. (1996). "The VFractal: A new estimator for fractal dimension of animal movement paths."

Landscape Ecology **11**(5): 289-297.

Nathan, R., W. M. Getz, et al. (2008). "A movement ecology paradigm for unifying organismal movement research." Proceedings of the National Academy of Sciences **105**(49): 19052-19059.

Pinaud, D. (2008). "Quantifying search effort of moving animals at several spatial scales using first-passage time analysis: effect of the structure of environment and tracking systems." Journal of Applied Ecology **45**(1): 91-99.

Polansky, L., G. Wittemyer, et al. (2010). "From moonlight to movement and synchronized randomness: Fourier and wavelet analyses of animal location time series data." Ecology **91**(5): 1506-1518.

Postlethwaite, C. M., P. Brown, et al. (2013). "A new multi-scale measure for analysing animal movement data." Journal of Theoretical Biology **317**(0): 175-185.

Rezek, I. A. and S. J. Roberts (1998). "Stochastic complexity measures for physiological signal analysis." Biomedical Engineering, IEEE Transactions on **45**(9): 1186-1191.

Riotte-Lambert, L., S. Benhamou, et al. (2013). "Periodicity analysis of movement recursions." Journal of Theoretical Biology **317**(0): 238-243.

Roberts, S., T. Guilford, et al. (2004). "Positional entropy during pigeon homing I: application of Bayesian latent state modelling." Journal of Theoretical Biology **227**(1): 39-50.

Sippel, T., J. Holdsworth, et al. (2011). "Investigating Behaviour and Population Dynamics of Striped Marlin (<italic>Kajikia audax</italic>) from the Southwest Pacific Ocean with Satellite Tags." Plos One **6**(6): e21087.

Thiebault, A. and Y. Tremblay (2013). "Splitting animal trajectories into fine-scale behaviorally consistent movement units: breaking points relate to external stimuli in a foraging seabird." Behavioral Ecology and Sociobiology **67**(6): 1013-1026.

Tomkiewicz, S. M., M. R. Fuller, et al. (2010). "Global positioning system and associated technologies in animal behaviour and ecological research." Philosophical Transactions of the Royal Society B: Biological Sciences **365**(1550): 2163-2176.

Wittemyer, G., L. Polansky, et al. (2008). "Disentangling the effects of forage, social rank, and risk on movement autocorrelation of elephants using Fourier and wavelet analyses." Proceedings of the National Academy of Sciences **105**(49): 19108-19113.